\documentstyle[11pt,paspconf,epsf]{article}

\begin{document}

\title{Star Formation in Polar-Ring Galaxies}
\author{Paul B.~Eskridge \& Richard W.~Pogge}
\affil{Department of Astronomy, The Ohio State University, \\Columbus, OH 43210, USA}

\begin{abstract}
Polar-ring galaxies are systems with two nearly orthogonal rotational axes, and
are therefore clearly the result of galaxy-galaxy interactions.  The host 
galaxies of polar rings are typically gas-poor early-type systems.  The rings
themselves, however, are typically gas-rich, blue, star-forming systems.  They
have traditionally been modelled as the remnants of accreted dwarf irregular
galaxies.  Published results on the molecular gas content, and nebular
abundances argue against this model.  We present H$\alpha$ imaging data for
three polar-ring galaxies, and derive their H\,II region luminosity functions.
In all three galaxies, the H\,II region luminosity functions are much steeper
than found for any dwarf galaxy.  The slopes are generally closer to those of
Sa galaxies.  Thus the H\,II region LFs of polar rings join the ISM abundances 
in resembling the very earliest-type spirals galaxies, rather than their
supposed dwarf irregular precursors.
\end{abstract}


\section{Overview of Observed Properties}

Polar-ring galaxies (PRGs) are typically early-type (E or S0) galaxies with 
luminous rings of stars, gas, and dust in nearly polar orbits (see Whitmore et
al.~1990 for a review).  Polar rings are
\begin{itemize}
\parsep 0ex
\item H\,I rich -- $M_{HI}$ up to $10^{10}M_{\odot}$ (Richter et al.~1994)
\item Blue -- $(B-V) \approx 0.5$ (Reshetnikov et al.~1994)
\item Dusty (Schweizer et al.~1983)
\item Actively star-forming (see \S 2).
\end{itemize}

In keeping with the above, the traditional model for the formation of PRGs 
involves the accretion of a gas-rich dwarf galaxy (a dI) by the host.  
Plausible alternative sources to dI galaxies are primordial H\,I, and material
captured from spiral disks.  Polar rings are typically much fainter than spiral 
disks (e.g.~Reshetnikov et al.~1994).  As a result, if spirals are the donors, 
only the outer parts of their disks have been captured. 

A problem with the dI accretion picture in particular is that the H\,I masses 
of polar rings can be quite large (Arnaboldi et al.~1997; Richter et al.~1994). 
While this is not a difficulty if the donor is primordial H\,I, or gas captured 
from spirals another problem remains:  All three of these sources have low 
heavy element abundances.  However, recent observations reveal that polar rings 
are often much stronger CO line-emission sources than are low-abundance 
systems (Galletta et al.~1997), and that the H\,II region abundances in the two 
PRGs studied to date are approximately Solar (Eskridge \& Pogge 1997, 1998).

\section{H$\alpha$ Imaging}

We are studying PRGs in order to understand their origin and evolution, and
their place in the context of galaxy evolution in general.  In order to 
evaluate the star-forming properties of polar-rings, we have been obtaining 
H$\alpha$ imaging of PRGs and PRG candidates.  Here, we report our results for
three kinematically confirmed PRGs:  NGC 2685, UGC 4385, and NGC 4650A.  Some 
properties of these galaxies are shown in Table 1.  A point worth emphasizing 
is that these are {\it not} intrinsically luminous galaxies; none are 
substantially brighter than M33 ($M_B = -18.4$).  Our absolute magnitudes are 
derived using the formalism of Yahil et al.~(1977), and adopting $H_\circ = 
75~{\rm km~s^{-1}~{Mpc}^{-1}}$.

{
\tolerance=500
 
\def\tabrule{\noalign{\hrule}}
\def\pz{\phantom{0}}
\def\pb{\phantom{-}}
 
\vskip0.2cm
\centerline{\bf Table 1 - Sample Properties}
\vskip0.3cm
\newbox\tablebox
\setbox\tablebox = \vbox {
 
\halign{#\pz\hfil&\hfil\pz#\hfil&\hfil\pz#\hfil&\hfil#\hfil&\hfil#\hfil&\hfil#
\hfil\cr
\tabrule
\noalign{\vskip0.1cm}
\tabrule
\noalign{\vskip0.2cm}
 
\ \ \ \ Name\ \  & $\rm B_{\circ}$ & Velocity & $\rm M_B$ & HI flux & 
$M_{HI}/L_B$ \cr
\ & \ & km/sec & \ & Jy km/sec & $M_{\odot}/L_{\odot}$ \cr
\noalign{\vskip0.1cm}
\tabrule
\noalign{\vskip0.3cm}
UGC 4385 & 14.4 & 1969 & -17.6 & 7.93$\pm$0.9 & 0.69 \cr
NGC 2685 & 11.9 & \ 876 & -18.5 & 34.2$\pm$4.0 & 0.32 \cr
NGC 4650A & 14.3 & 2910 & -18.4 & 23.2$\pm$2.4 & 1.09 \cr
\noalign{\vskip0.2cm}
\tabrule
\noalign{\vskip0.2cm}
}
}
\centerline{ \box\tablebox}
}

\begin{figure}
\plotone{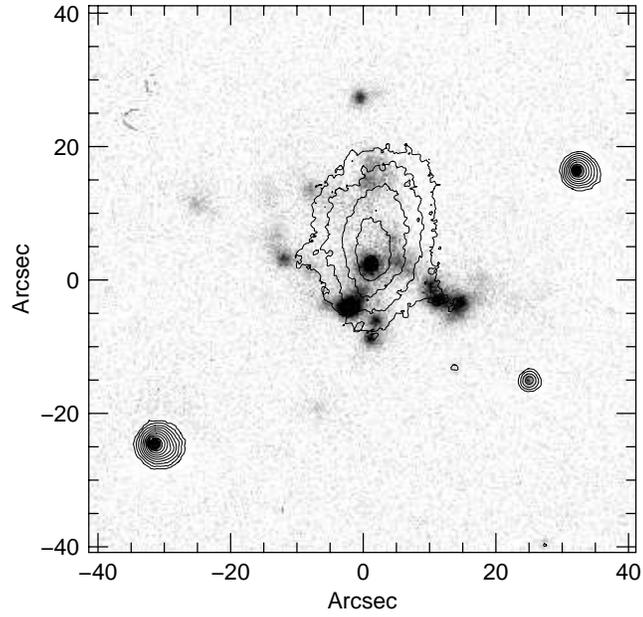}
\caption{UGC 4385.  The greyscale is H$\alpha + [N\,II]$ emission.  The 
contours are red continuum.}
\end{figure}
\begin{figure}
\plotone{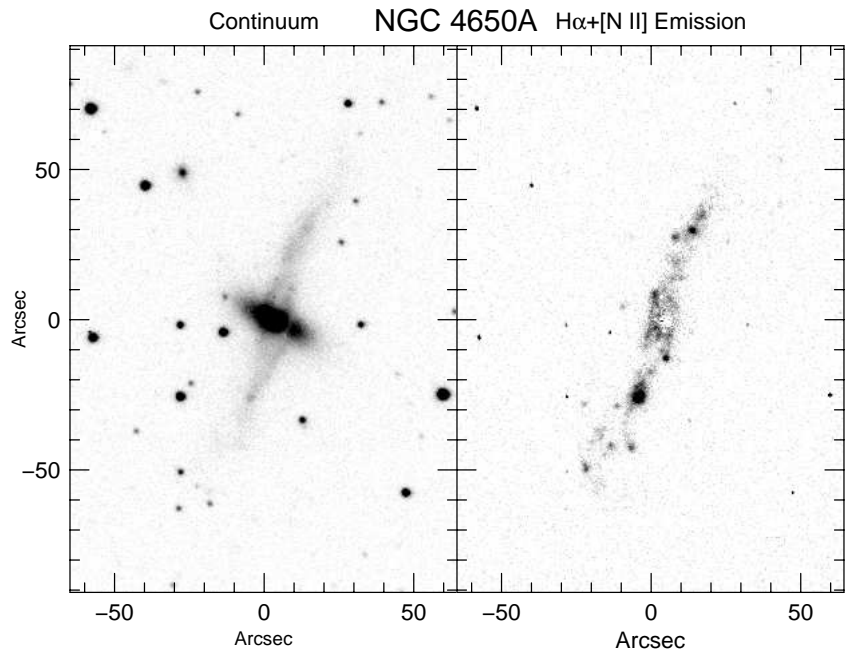}
\caption{NGC 4650A.  The red continuum is shown on the left.  The H$\alpha + 
[N\,II]$ emission is shown on the right.}
\end{figure}

In addition to having rich populations of H\,II regions, these three systems
form an interesting set in that NGC 2685 is the prototype of PRGs with rings
that are small compared to the host galaxy, whereas NGC 4650A is the prototype
of the large-ringed PRGs.  UGC 4385 is an excellent candidate for a forming
PRG, as it has a very disturbed morphology, but well-behaved kinematics 
(Reshetnikov \& Combes 1994).  In Figure 1, we show an H$\alpha + [N\,II]$ 
emission-line image of UGC 4385.  The contours are the red continuum light.  
Figure 2 shows emission-line and continuum images of NGC 4650A.  Our imaging 
data for NGC 2685 can be found in Eskridge \& Pogge (1997).

\section{H\,II Region Luminosity Functions}

From the H$\alpha$ images, we have derived H\,II region luminosity functions 
(LFs) for our three targets.  These are shown in Figure 3.  We note that the 
data for UGC 4385 and NGC 4650A are not photometrically calibrated.  We have 
fit these LFs with power laws, following the spiral galaxy study of Kennicutt 
et al.~(1989).  In all cases, the bright-end slope of the LF is steep.  For NGC 
2685 and UGC 4385 (top and middle panels of Figure 3), we find $\alpha \approx 
-2.5$, and for NGC 4650A (bottom panel of Figure 3), $\alpha \approx -3$.  
These are plotted as solid lines in Figure 3.  

\begin{figure}
\plotone{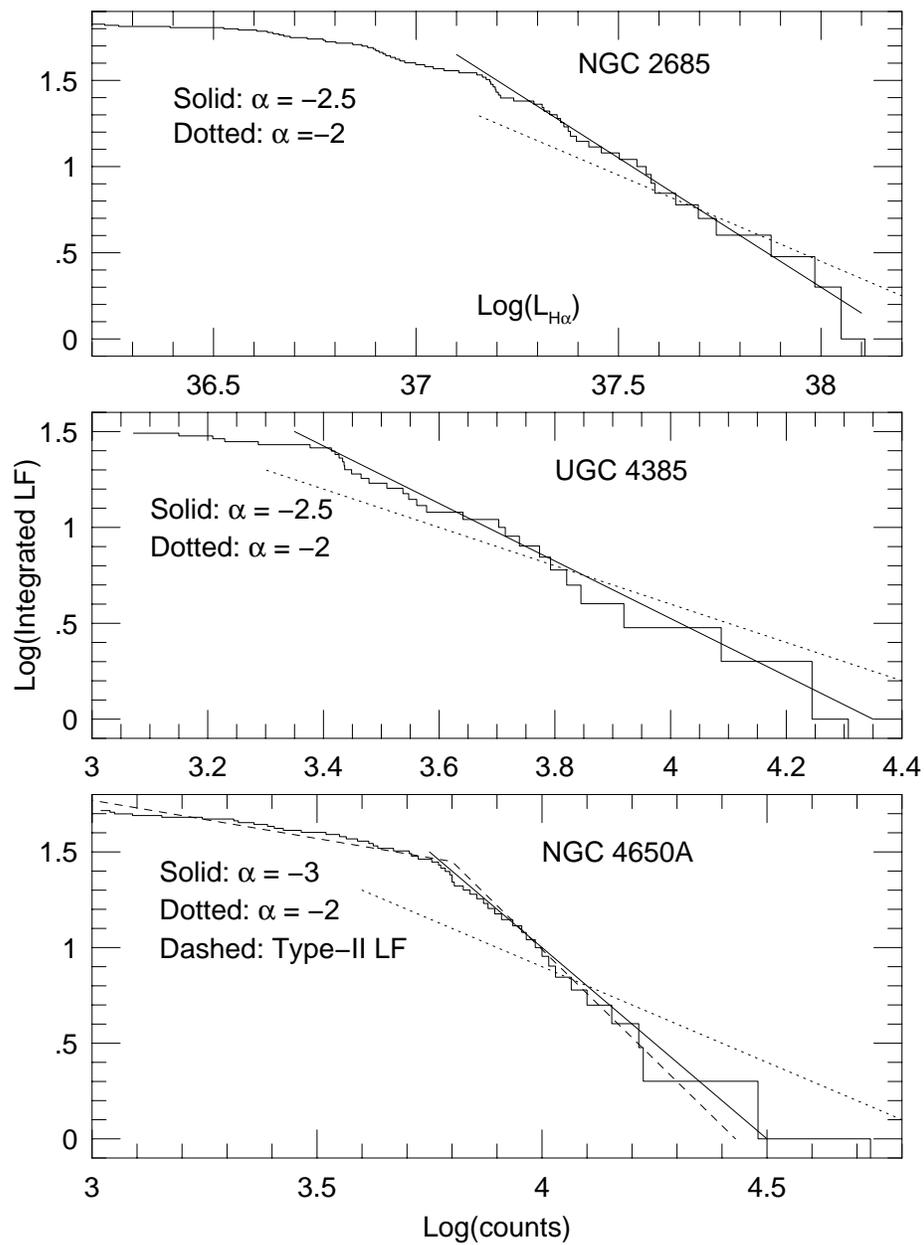}
\caption{Integrated H\,II region LFs for the three target galaxies.}
\end{figure}

As a comparison, typical Sb and Sc spirals have $\alpha \approx -2 \pm 0.3$ 
(Kennicutt et al.~1989).  The dotted lines in Figure 3 show how poorly the 
canonical $\alpha = -2$ power law fits our data.  Irregular galaxies, the 
putative sources of the polar-ring material, have $\alpha \ga -1.8$ (Kennicutt 
et al.~1989; Banfi et al.~1993).  Sa spirals tend to have steeper LFs, with 
$\alpha \approx -2.6$ (Caldwell et al.~1991).  Our results show that polar 
rings have H\,II region LFs that decline much more steeply than any dI studied 
to date.  Indeed, the slopes we find are steep even in comparison to typical Sb 
and Sc spirals.  They are instead more typical of what is seen in Sa spirals.

We note that the shape of the LF for NGC 4650A is similar to the type II LF
identified by Kennicutt et al.~(1989).  This is shown by the dashed line in
Figure 3c.  Type II LFs have a very steep upper end ($\alpha \approx -3.3$), 
with a break to a shallower slope ($\alpha \approx -1.4$) at fainter 
luminosities.  As noted above, our data for NGC 4650A are not photometrically 
calibrated, so we do not know if the break we see is at the correct luminosity. 
It could, instead be due to sample incompleteness.

\section{Summary}

PRGs are a class of interacting galaxies wherein the material of the precursors 
are still distinguishable.  Formation models favor low initial heavy element
abundances, but both molecular and optical line observations indicate high 
abundances.  Many polar rings are rich in H\,II regions.  This allows for 
statistical study of the ensemble of H\,II regions in individual systems.  We 
have studied the H\,II region luminosity functions for the three PRGs NGC 2685, 
UGC 4385, and NGC 4650A.  We find all three to have very steep slopes for their 
LFs ($-3 \leq \alpha \leq -2.5$).  That is, there is a deficit of bright H\,II 
regions in polar rings compared to what is seen in dIs and mid- to late-type
spirals.  Thus, both the abundances and the H\,II region LFs of polar rings are 
very different than those of dIs.  Instead, both appear more similar to what is 
found in $\sim$Sa spirals.  This argues strongly that polar rings are not the 
remains of accreted dI galaxies. 

Leaving aside the issue of abundances, we conclude with some speculation 
on the following question:  Why do polar rings have such steep H\,II region 
LFs?

It could be an IMF effect:  Polar rings could make fewer very high mass stars.
However, recent work on H\,II region luminosity functions in spirals indicates
that this is not typically a good explanation for their LF slopes 
(e.g.~Bresolin \& Kennicutt 1997; Oey \& Clarke 1998).  More plausibly, it may 
be a cluster mass function effect:  Massive cluster formation is inhibited in
the polar ring environment.  One plausible mechanism for this is strong shear
in the polar ring gas, although this does not appear to be the case for NGC
4650A (Arnaboldi et al.~1997).  Another possibility is that it may be an
abundance effect.  Detailed kinematic and abundance studies are needed to 
explore this question further.

\acknowledgments

We are happy to thank the LOC for a wonderful and intimate meeting, and for
their excellent hospitality.  PBEs travel was supported by NASA grant NAG 
5-2990.

\end{document}